\documentclass[fleqn,12pt,twoside]{article}
\usepackage{espcrc1}

\usepackage{graphicx}

\title{{\bf Electroproduction of K$^{*0}$ mesons at CLAS} }

\author{I. Hleiqawi\address[OU]{Department of Physics and Astronomy,
        Ohio University, \\ 
        Athens, OH 45701}%
        \thanks{For the CLAS collaboration.}
        }
       
\begin{document}

% typeset front matter
\maketitle
%%%%%%%%%%%%%%%%%%%%%%%%%%%
\begin{abstract}
%%%%%%%%%%%%%%%%%%%%%%%%%%%%%
The electroproduction of K$^{*0}$ mesons using the CLAS detector is 
described. Data for two electron beam energies, 4.056 and 4.247 GeV, 
were measured and the normalized yields are compared.
\end{abstract}
%%%%%%%%%%%%%%%%%%%%%%%%%%%%%%%%%
\section{INTRODUCTION}
%%%%%%%%%%%%%%%%%%%%%%%%%%%%%%%%%
Quark models predict baryon resonances \cite{isgur1} that 
have not yet been observed via strong interactions. These resonances
could be detected via electro- and photoproduction of strange mesons.
Of special interest are the nucleon resonances N$^*$ which 
some of them could couple strongly to $K\Lambda$ and 
$K\Sigma$ \cite{capstick2}. 
Moreover, higher mass nucleon resonances could favor decaying
into $K^*\Sigma$, near threshold. Using a quark-pair creation 
model \cite{capstick2}, a study of N$^*\rightarrow K^*Y$, where $Y$ is 
the hyperon, shows 
that most $K^*Y$ decay branching ratios are small due to the high 
thresholds of these channels. Only a few low-lying negative-parity
states are predicted to be strongly coupled to $K^*\Sigma$ channels, 
e.g. N(2070), $\Delta$(2140), $\Delta$(2145). 
In addition, vector meson production could be used to investigate 
below-threshold resonance couplings to $K^*$Y.
           
Experimental studies of neutral vector mesons used to concentrate 
on high energy regions that are dominated by the diffractive process 
and which could be accounted for by a soft Pomeron exchange 
model. Only recently, the study of non-diffractive mechanism in
the vector meson production, near threshold, has become possible 
via resonance excitations. Vector meson electro- and
photoproduction near threshold might provide good knowledge 
about these resonances, their internal structure, and their couplings 
to vector mesons. This has been the main motivation for doing experiments 
to study strange mesons electroproduction off the 
proton, $ep\rightarrow e'KY$,
\begin{equation}
\label{Eq:electro}
~~~~~~~~~~~~~~~{\rm ep} \rightarrow {\rm e'K}^{+}\Lambda~~~~~~~{\rm ep} \rightarrow 
{\rm e'K}^+\Sigma{^0}~~~~~~~{\rm ep} \rightarrow {\rm e'K}^{{\rm *0}}\Sigma{^+}
\end{equation}  

Much work and publications have been done on the first two channels
\cite{Feuerb}.
However, the third channel has barely been studied, because of its 
small cross section and the difficulty of detecting the $K^*$ decay.
The availability of the high 
intensity electron facility and the Cebaf Large Acceptance Spectrometer 
(CLAS) in Hall B \cite{clas} at JLAB, and other facilities, has made it 
possible to study this channel and opened new avenues to search for 
``missing resonances". Here, $K^{*0}$ electroproduction results are 
presented. Preliminary results of this reaction are also shown in 
the thesis of Ref. \cite{weisberg}. 
\section{THEORETICAL BACKGROUND}
%%%%%%%%%%%%%%%%%%%%%%%%%%%
A theoretical model in which a quark model, 
with an effective Lagrangian, approach to vector meson 
production \cite{zhao3}, near threshold, has been developed for
K$^{*0}$ production \cite{zhao}. It is the first theoretical attempt 
to study nucleon resonances and to present quark model predictions 
for the K$^{*0}$ production. In addition to using
common quark model parameters, this model (i) uses two free parameters:
the vector and tensor couplings for the quark-K$^*$ interaction.
They are the basic parameters in this model and are related to the 
K$^*\Sigma$N$^*$ couplings that appear in the quark model symmetry limit,
and (ii) adopted the SU(3)-flavor-blind assumption of non-perturbative QCD,
which suggests the above two parameters should have values close
to those used in the $\omega$ and $\rho$ meson photoproductions. 
Our experimental data should play a rule in tuning these parameters
as well as testing and improving this model.
%{\underline {Included resonances}}: 
%(1) low-laying resonances (n $\leq$ 2 harmonic oscillator shells)
%    are included explicitly in the formalism and 
%(2) higher mass
%    resonances (n $>$ 2) are treated as being degenerate by summing 
%    over all states for each n.

The production of K$^*$ vector mesons is related to other 
strangeness productions, Eq.~(\ref{Eq:electro}), as well as 
non-strange vector meson production. At the {\it hadronic level}, 
these reactions are related
to each other since one reaction contains the meson production in the 
other one as the t-channel exchanged particle, and therefore constraining
the range of available couplings. This allows K$^*$ and K production to
use the same observables, which are obtained from non-strange vector
meson production, $\rho$ and $\omega$ in the resonance region 
\cite{zhao3}. At the {\it quark level}, 
both K and K$^*$ productions involve the creation of s$\bar{s}$ 
pair production, from the vacuum, in the SU(3) quark model. 
In this model, the s quark couples to the meson in the same way as 
the u and d quarks, assuming quarks have the same masses, i.e. 
assumption (ii) above.

\section{EXPERIMENT}
%%%%%%%%%%%%%%%%%%%%%%%%%%%%%%%%%%%
The K$^{*0}$ electroproduction data were extracted from``e1b"
data set using the CLAS detector \cite{clas}, at Jefferson 
Lab's Hall B. CLAS  consists of:
(1) the main torus: six superconducting magnetic coils making 
regions, or ``sectors". Their toroidal magnetic 
field deflects charged particles toward or away from the beam 
line, while keeping the azimuthal angle unchanged,
(2) a forward-angle electromagnetic calorimeter (EC): 
it is located in the forward region of each CLAS sector
and covers up to 45$^{\circ}$ of the polar angle. 
It detects particles moving forward and distinguishes 
electrons from pions,
(3) Cherenkov Counter (CC): covers the same angular range above.
It is used along with the calorimeter to create a
coincidence trigger,
(4) three Drift Chambers (DC): located in each sector and they 
determine the trajectories
of charged particles from which their momenta are reconstructed.
They cover about 80 $\%$ of the azimuthal angle
and a polar range from 8$^{\circ}$ to 142$^{\circ}$, and
(5) Scintillator Counters (SC): an array of 288 scintillator 
counters where the charged particles times of flight and their
energies are determined. They cover the same angular 
range as the DC.

In this analysis we took data at 4.056 
and 4.247 GeV electron beam energies, with a liquid hydrogen
target of length 5 cm.
All data (430 and 610 million triggers, respectively)
were taken in Feb. 1999.
The toroidal coils current was 2250 A, corresponding to a magnetic 
field at 60\% of the maximal field setting.

%%%%%%%%%%%%%%%%%%%%%%%%%%%%%%%%%%%%%%%%%%%%%%%%%%%%%%%%%%
We binned the K$^{*0}$ electroproduction data in the c.m. 
energy W, from 2.1 to 2.5 GeV, 
and the 4-momentum transfer Q$^{\rm 2}$, from 0.75 to 1.5 (GeV/c)$^2$,
which are determined entirely by the electron kinematics.
Due to low statistics, the data were binned into large bins: 
(i) 100 MeV in W, 
(ii) integrated over the above broad range in Q$^{\rm 2}$, and 
(iii) integrated over the full angular range of $\theta$ and $\phi$.
%%%%%%%%%%%%%%%%%%%%%%%%%%%%%%%%%%%%%%%%%%%
%%%%%%%%%%%%%%%%%%%%%%%%%%
\begin{figure}[htb]
\begin{center}
%\framebox[79mm]{\rule[-26mm]{0mm}{52mm}}
\vspace{13.5mm}
\centering{\includegraphics{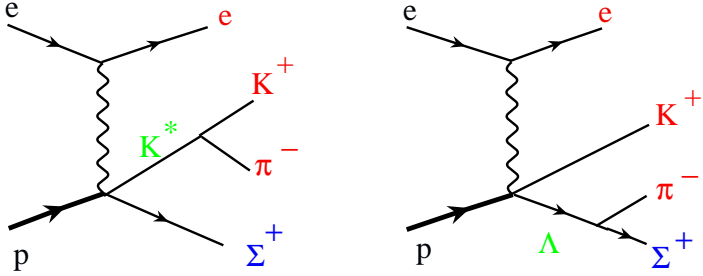}}
\end{center}
\caption{{\footnotesize{Feynman diagrams of the reaction of interest (left) and
  one of the physics background channels, e.g. $\Lambda$(1520) production,
  Eq.~(\ref{Eq:our2}). The detected particles are $e'K^+\pi^-$. 
         }}}
\label{fig:fig1_ks_lam}
\end{figure}
%%%%%%%%%%%%%%%%%%%%%%%%%%%%%%%%%%%%%%%%%%%%

In addition to the reaction of interest, $ep \to e'K^{*0}\Sigma^+$,
where K$^{*0}$(892) decays immediately into two detected 
particles, $K^{*0} \to K^+\pi^-$, there are other physics 
background contributions, from K$^+$ production,
\begin{equation}
\label{Eq:our2}
~~~~~~~~~~~~~~~~~ ep\rightarrow  e'K^+ \Lambda (1520)
\end{equation}  
\begin{equation}
\label{Eq:our3}
~~~~~~~~~~~~~~~~~ ep \rightarrow  e'K^+ Y^* 
\end{equation}

\noindent   
where the $\Lambda$(1520), Eq.~(\ref{Eq:our2}), decays 
with 14\% probability
to $ \Lambda (1520) \rightarrow \pi^-\Sigma^+$. 
That is, Eqs.~(\ref{Eq:our2} and \ref{Eq:our3}) 
are background contributions to our final state, $K^+\pi^-\Sigma{^+}$.
Eq.~(\ref{Eq:our3}) represents physics background from 
other excited states of $\Lambda$ family, e.g. $\Lambda$(1600),
$\Lambda$(1670), $\Lambda$(1690). These resonances have 3- or 4-star
ratings as well as several exited states above 1.8 GeV.
Diagrams of the K$^{*0}$(892) production (signal)
and the $\Lambda$(1520) production 
are shown in Fig.~(\ref{fig:fig1_ks_lam}).
%%%%%%%%%%%%%%%%%%%%%%%%%%

The detected particles are $e$, $K^{+}$, and $\pi^-$, while
K$^{*0}$ and $\Sigma^+$ are undetected directly. 
The K$^{*0}$ was identified from the invariant mass of the K$^+\pi^-$
system, which peaks at 0.892 GeV, while $\Sigma^+$,
from which we obtained the yields, was identified by applying 
cuts on both K$^+$ mass and the missing mass of K$^{*0}$, 
as explained below. 

We have two major background sources: 
(i) from the $\pi^+\pi^-$ events, where the $\pi^+$ is mis-identified 
    as a K$^+$, and 
(ii) physics backgrounds from other channels, Eq.~(\ref{Eq:our3}). 
     To remove both
     background contributions and to identify $\Sigma^+$, we used 
    the side-band technique and applied the 
    following mass cuts on both K$^+$ and K$^{*0}$ mass distributions: 
    (1) a cut on the K$^+$ mass peak and a side-band cut, 
    (2) two similar cuts on the K$^{*0}$ invariant mass distribution. 
        The side-band cuts on both K$^+$ and K$^{*0}$ produce the 
        pion background and the physics background, respectively.
Combinations of the above four cuts produce two $\Sigma^+$ peaks shown 
in Fig.~(2), 
       (i) top (K$^{*0}$ cut): cuts on both K$^+$ and K$^{*0}$ peaks
         give the signal (upper/green), and
       (ii) bottom (Y$^{*0}$ cut): K$^+$ peak along with the K$^{*0}$
        side-band cuts give the physics background contribution (lower/red).
The other two combinations give the pion background shown, in yellow, 
under each peak. 
In particular, the background from $\Lambda$(1520) production,
Eq.~(\ref{Eq:our2}), was removed by applying a cut on the K$^+$ missing
mass.

\newpage
%%%%%%%%%%%%%%%%%%%%%%%%%%%%%%%%%%%%%%%%%%%%%%%%%%%%%%%%%%%%%%%%%%%%%
\begin{figure}[h]
\vspace{58mm}
\centering{\includegraphics{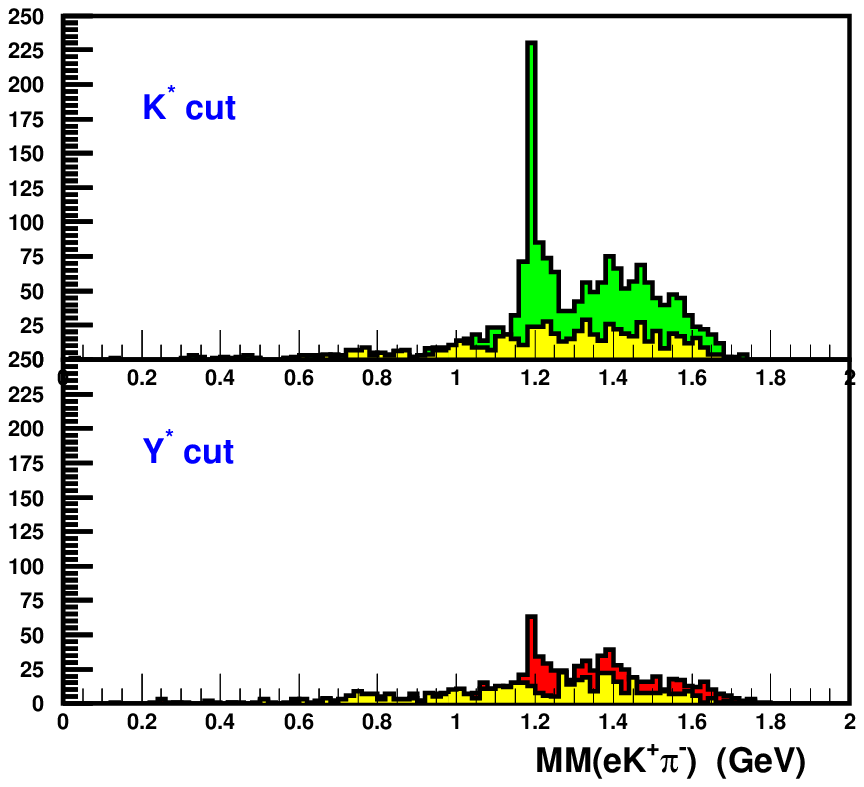}}
%\caption{\footnotesize{ The final $\Sigma^+$.} }
\label{fig:fig2_sigpeak}
\end{figure}
%%%%%%%%%%%%%%%%%%%%%%%%%%%%%%%%%%%%%%%%%%%%%%%%
\begin{figure}[h]
\vspace{-21.mm}
\centering{\includegraphics{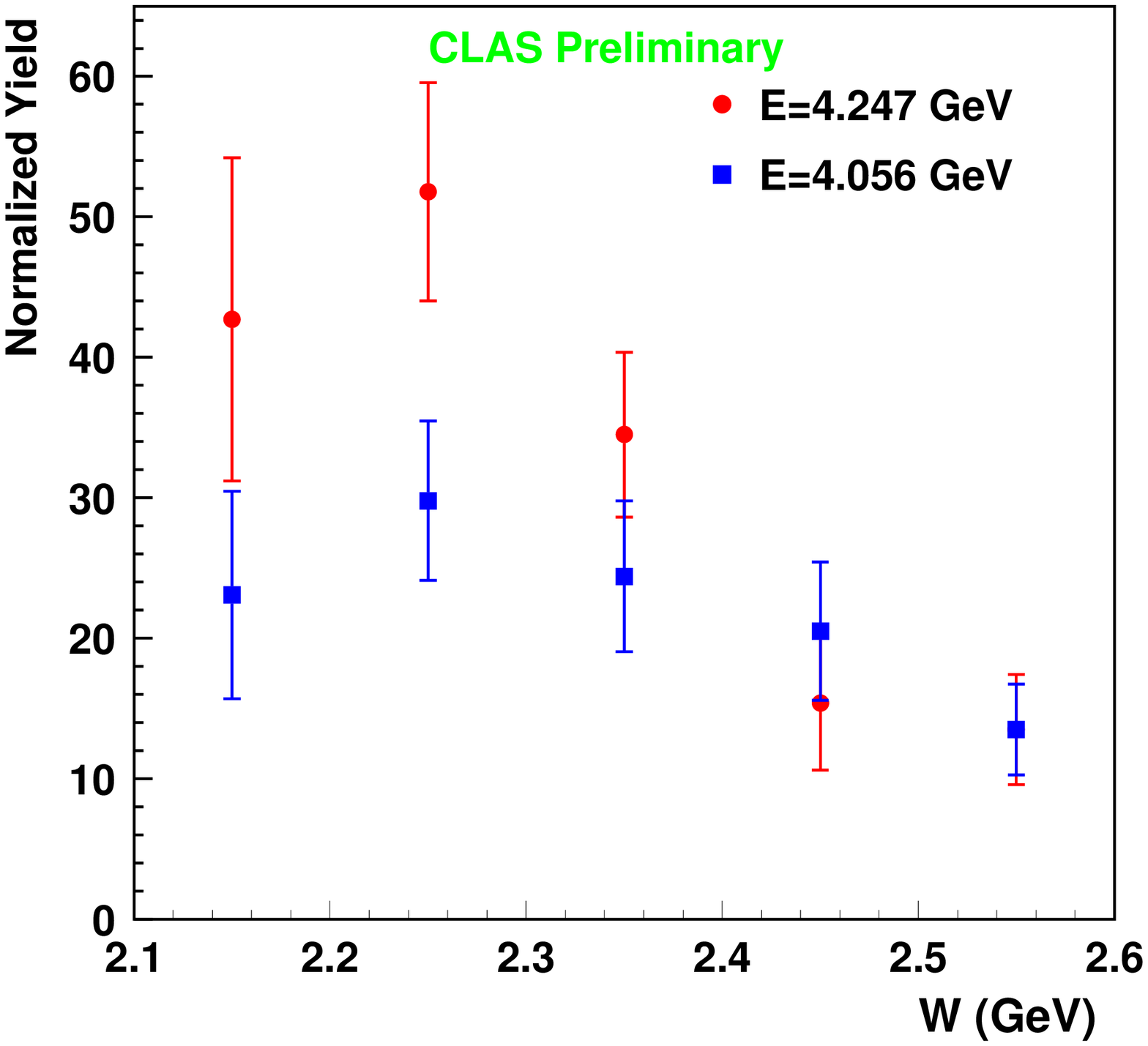}}
%         {   \caption{Normalized fields (arbitrary units).}  }
\label{fig:fig3_norm_yeild}
\end{figure}
%%%%%%%%%%%%%%%%%%%%%%%%
\vspace{-6.5mm}
~Figure 2. {\footnotesize{K$^{^*0}$ missing mass, $\Sigma^+$ peak.}}~~~~~~~~~~~~~~~~~
Figure 3. {\footnotesize{Normalized yields (arbitrary units)}}
%%%%%%%%%%%%%%%%%%%%%%%%%%%%%%%%%%%%%%%%%%%%%%%%% 

%vspace{3mm}    
%%%%%%%%%%%%%%%%%%%%%%%%%%%%%%%%%%%%
\section{RESULTS AND CONCLUSIONS}
%%%%%%%%%%%%%%%%%%%%%%%%%%%%%%%%%%%%
We subtracted the pion background events from each
$\Sigma^+$ peak, Fig.~(2), obtaining two yields, 
one from K$^*$ cut and the other
from Y$^*$ cut. Subtracting the latter yields from the 
former one resulted in the final yields.
The yields were then corrected for the CLAS detector acceptance and
normalized by the virtual photon flux. 

The measurements have been finished and analysis of the data is in 
progress. Preliminary values of the normalized yields are shown in 
Fig.~(\ref{fig:fig3_norm_yeild})
as a function of the center-of-mass energy W. Currently, these 
data are being compared with the theoretical model (not shown).
Our preliminary results show disagreement between the two beam energies,
and further work is needed. However, the disagreement could be due to:
(i) broad $Q^{\rm 2}$ bin: a difference between the two $Q^{\rm 2}$ 
    averages, at a
    particular W value, due to our broad $Q^{\rm 2}$ bin,
    might have contributed to this disagreement, 
(ii) $\epsilon$ dependence: the virtual photon flux dependence on 
     the polarization constant, $\epsilon$, which have different values
     for the two beam energies, might have contributed to this
     disagreement too,
(iii) acceptance: the model from which we calculated the acceptances 
     is still under improvements, Sec. 2. Improving the model and 
     tuning 
     its free parameters are expected to improve the acceptances, and 
     therefore narrowing the difference between the normalized yields. 
     Improving the model and calculating the final cross sections are 
     in progress.
     
\vspace{5mm}
{\it{ I would like to thank Ken Hicks and Avto Tkabladze for
their help in this project, and also to Stepan Stepanyan
for valuable discussions.}}

\end{document}